\documentclass[preprint,showpacs,preprintnumbers,amsmath,amssymb]{revtex4}

\usepackage{graphicx}% Include figure files
\usepackage{dcolumn}% Align table columns on decimal point
\usepackage{bm}% bold math

\begin{document}

\preprint{APS/123-QED}

\title{The effect of spatially correlated noise on coherence resonance in a network of excitable cells}

\author{Okyu Kwon}
\email{okyou@kaist.ac.kr}
\affiliation{Department of Physics, Korea Advanced Institute of Science and Technology, Daejeon 305-701, Korea}

\author{Hang-Hyun Jo}
\affiliation{Department of Physics, Korea Advanced Institute of Science and Technology, Daejeon 305-701, Korea}

\author{Hie-Tae Moon}
\affiliation{Department of Physics, Korea Advanced Institute of Science and Technology, Daejeon 305-701, Korea}

\date{\today}

\begin{abstract}
We study the effect of spatially correlated noise on coherence resonance (CR)
in a Watts-Strogatz small-world network of Fitz Hugh-Nagumo neurons, where
the noise correlation decays exponentially with distance between neurons.
It is found that CR is considerably improved just by a small fraction of
long-range connections for an intermediate coupling strength. For other
coupling strengths, an abrupt change in CR occurs following the drastic
fracture of the clustered structures in the network. Our study shows that
spatially correlated noise plays a significant role in the phenomenon of CR
through enforcing the clustering of the network.
\end{abstract}

%05.40.-a: Fluctuation phenomena, random processes, noise, and Browinan motion
%05.45.-a: Nonlinear dynamics and nonlinear dynamical systems
%05.45.Xt: Synchronization; coupled oscillators
%87.10.+e: General theory and mathmatical aspects
%87.18.Sn: Neural networks
%87.18.Bb: Computer simulation

\maketitle

%Introduction
The response of nonlinear systems to noise has attracted large attention.
Especially stochastic resonance (SR) has been studied very extensively during
the last two decades due to a number of applications in many fields, from
physical to biological systems \cite{srBenzi,srGammaitoni,srWiesenfeld,srBio,
srBio2}. The main result of SR, which is somewhat counterintuitive,
shows that noise at a proper strength optimizes the response of a nonlinear
system to a subthreshold periodic signal. An optimal strength of noise can
induce the most coherent motion in the system. SR-like behavior of the coherent
motion can be induced purely by noise even in the absence of an external
periodic signal for a system at a saddle-node bifurcation point \cite{crGang}.
This phenomenon has been called coherence resonance (CR) or autonomous SR.
In general, various excitable systems such as the Fitz Hugh-Nagumo model
\cite{crFHN}, the Plant model, the Hindermarsh-Rose model \cite{crHR} and the
Hodgkin-Huxley model \cite{crHH} exhibit such noise induced coherent motion.

Recently, SR and CR in coupled or extended systems have become an
interesting issue, and some new features have been demonstrated, namely the
noise-enhanced phase synchronization \cite{nisync}, the noise-induced
spatiotemporal pattern formation \cite{stsr} and the noise-enhanced wave
propagation \cite{nep,nep2}. Also, the phenomena called array enhanced
stochastic resonance (AESR) and array enhanced coherence resonance (AECR)
\cite{aesrLindner,aecrHu,aecrZhou,aesrGao,aecrOkyu} have drawn interests among
researchers in recent years. It is now understood that in spatially extended
systems (i) the topology of connecting structure and (ii) noise correlation
among the elements are the significant ingredients on the collective behavior
of the systems.
Actually, the connection topology of a variety of extended systems can be
described by complex networks \cite{compnet}.
Especially many biological neural networks present clear clustered structure
and sparsely long-range random connectivity \cite{bioneuralnet}.
On the other hand, spatially correlated noise has been considered relevant
for biological systems. However, most previous studies have not dealt with
these factors together \cite{aecrZhou,aesrGao,aecrOkyu,crcnWang}.
In this study, we plan to add the element of spatially correlated noise in the
system of excitable cells and to study the corresponding dynamics of CR.

%Model Description
As a model, we consider a system of coupled excitable cells on ``small-world"
network, introduced by Watts and Strogatz \cite{smwldnet} in the presence of
a spatially correlated noise. Each cell is a Fitz Hugh-Nagumo (FHN) neuron
which is a simple but representative model of excitable neuron \cite{fhn}.
$N$ neurons in a ring lattice are diffusively coupled as following:
\begin{eqnarray}\label{eq1}
\epsilon \dot{x_i} = x_i - \frac{x_i^3}{3} - y_i + \sum_j g_{ij} ( x_j - x_i ), \\
\dot{y_i} = x_i + a + \xi_i,
\end{eqnarray}
where $x_i$ is the fast voltage variable and $y_i$ is the slow recovery
variable of $i$th neuron. $\epsilon$ and $a$ are a time scale and a bifurcation
parameter respectively. We fix $\epsilon = 0.01$ and $a=1.03$ for all $N=101$
neurons. For $|a|>1$, a single FHN neuron has only a stable fixed point.
While for $|a|<1$, a limit cycle occurs. $g_{ij}$ is a coupling strength
between two neurons $i$ and $j$. If connected, they have the coupling strength
$g_{ij}=g$, otherwise $g_{ij}=0$. The connectivity pattern can vary with
parameter $p$, which measures the network randomness.
$\xi_i$ is spatially correlated noise with intensity $D$.

The spatially correlated noise $\xi$ is generated by summing $N$ Gaussian white
noises $\zeta$ with correlation function $C$.
\begin{eqnarray}\label{eq2}
\xi_i = \frac{1}{\sqrt{\sum_{k\in\Lambda} C_k^2}} \sum_{k\in\Lambda} \zeta_{i+k} C_k \\
\Lambda = \{ -4\lambda,\dots,-2,-1,0,1,2,\dots,4\lambda \},
\end{eqnarray}
where $\zeta_i$ is a Gaussian white noise with zero mean and correlation given
by $\langle \zeta_i(t) \zeta_j(t') \rangle = D \delta_{ij} \delta(t-t')$; $D$
denoting the noise intensity. And the correlation function is defined by
$C_k = \exp(-2 k^2/ \lambda^2)$. According to the above method, we can get
the spatially correlated noise which obeys the correlation function with
decay constant $\lambda$ as following:
\begin{equation}\label{eq3}
\langle \xi_i(t) \xi_j(t') \rangle = D \exp \left(
-\frac{|i-j|^2}{\lambda^2} \right) \delta (t-t'),
\end{equation}
where $i$ and $j$ denote the spatial positions of neurons in the ring lattice.
Consequently $|i-j|$ represents a distance along the ring lattice not the
connection topology. The systems are numerically integrated by the method
of Fox {\it et al.} \cite{numericalFox} with the time step
$\Delta t = 0.002$ {\it t.u.} (time units).

As a quantitative observable of a neuron showing CR, a temporal coherence is
calculated by
\begin{equation}\label{eq4}
S_i = \frac{\langle \tau_i \rangle_t}{\sqrt{\textrm{Var}(\tau_i)}}
\qquad i = 1,2, \dots ,N,
\end{equation}
where $\tau_i$ is the ensemble of time interval of inter spike and $S_i$
denotes temporal coherence factor of $i$th neuron. Here $\langle\cdot\rangle_t$
denote average over time. The coherence factor $S$ of the system is computed
by averaging $S_i$ over all $N$ neurons. A larger $S$ implies that the
inter spike intervals of neurons are more uniform.

\begin{figure}
\includegraphics{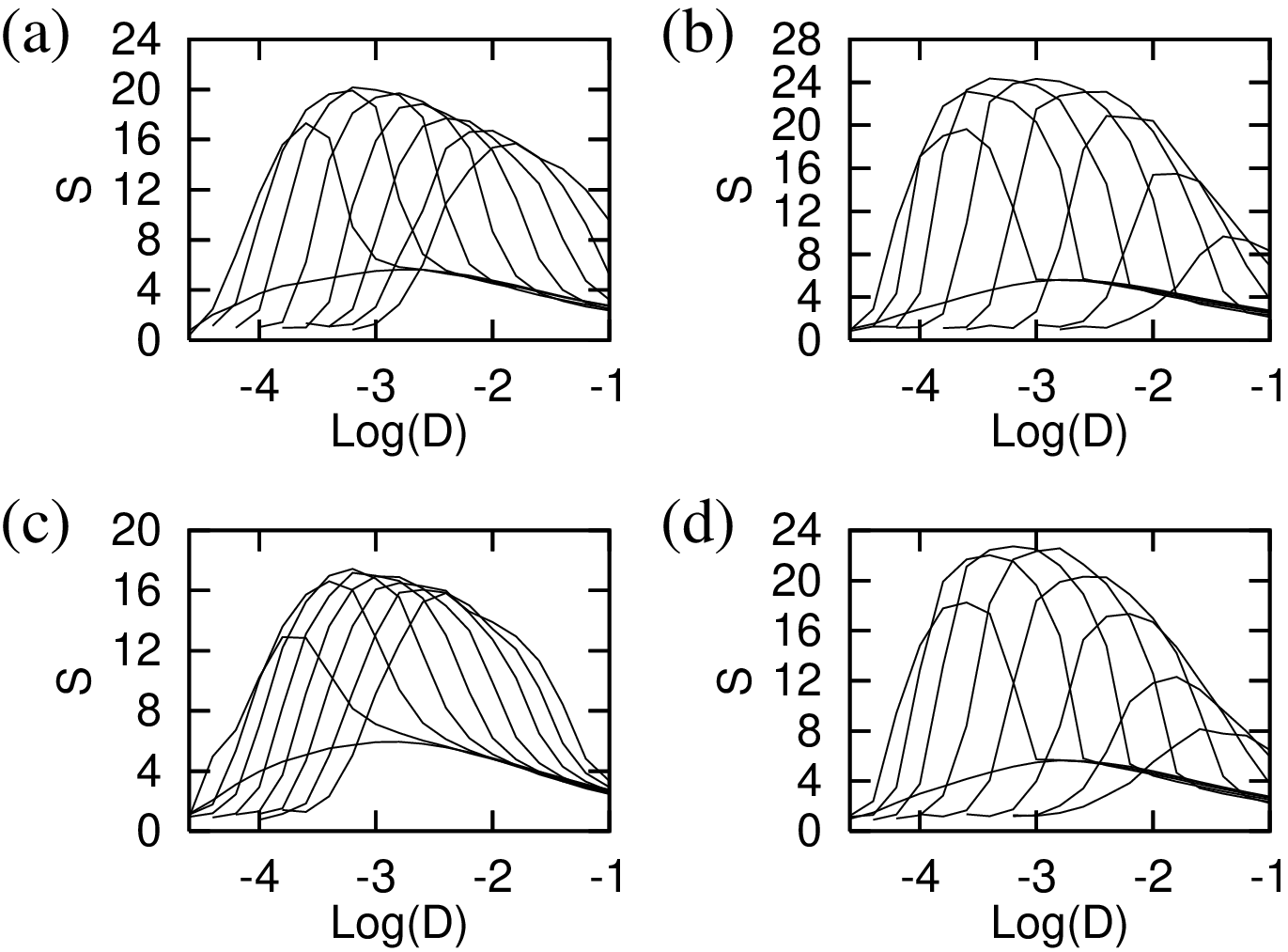}
\caption{\label{fig1} The coherence factor $S$ versus noise intensity $D$ for
several values of coupling strength on (a) regular network ($p=0$) and (b)
completely random network ($p=1$) with spatial correlation length of noise
$\lambda=0$. (c), (d) are the same curves for $p=0$ and $p=1$ with $\lambda=2$
respectively. Average number of neighbors $k=6$. Number of elements $N=101$.
Coupling strength $g$ varies from $g=10^{-2.5}$ to $g=10^{-0.5}$ with 0.25 step
of exponent. Base curves represent for $g=10^{-2.5}$.
As coupling strength increases, the peak of $S$ shifts to a stronger noise.}
\end{figure}

%Numerical results
A general feature of SR and CR is that there exists an optimal noise
intensity at which the coherence factor is maximized. In the cases of AESR and
AECR there exists an optimal coupling strength additionally \cite{aesrLindner,
aecrHu,aesrGao}. Our numerical simulations verify the above results
(see Fig. \ref{fig1}), that is, both the optimal noise intensity and the
optimal coupling strength exist, moreover, regardless to the network
randomness $p$ and the correlation length $\lambda$ of the noise. For a very
weak coupling $g\approx10^{-2.5}$, each element is effectively independent.
Therefore a coherence factor curve has the similar appearance to that of a
single uncoupled neuron with the same parameters. As coupling strength
increases, the impact of noise-induced firing events propagate through the
connection topology. This cooperation of noise-induced individual and
coupling-induced mutual excitation enhances a coherent motion in the coupled
system. However, a strong coupling rather disturbs excitation of elements
because those elements that are excited by noise are strongly attracted to a
resting state by resting neighbors. For this reason, a stronger noise is needed
to overcoming stronger coupling and to excite the neurons. In this case a
global synchronization emerges due to strong coupling while the temporal
coherence of the system is somewhat reduced due to strong noise. As seen in
Fig. \ref{fig1}, the resonance curve shifts to right side gradually and its
peak rises up first and then drops down according to the coupling strength.

How do the connection topology and the spatial correlation of noise influence
the coupled excitable neurons? To focus on the question the effect of
structural changes of connectivity on the spatial correlation of noise must
be explained in advance. For a regular network ($p=0$), if $\lambda$ is large
enough, each neuron interacts only with the ones exposed to the correlated
noise. As $p$ increases each neuron is able to interact with a distant one 
which is exposed to an uncorrelated noise. Consequently for a proper length of
$\lambda$, increasing of $p$ reduces the correlation of the noise and
simultaneously increases the small-world effect between coupled neurons.
However, when $\lambda=0$ every neuron is already exposed to a totally
uncorrelated noise, i.e. local noise, so varying $p$ does not alter the
correlation of noise. In this case the variation of $p$ influences the
small-world effect only.

Now let us figure out the peculiar result according to $p$ and $\lambda$.
In Fig. \ref{fig1} the maximum value of each curve is called a maximal
coherence factor $S_m$ and the noise intensity at $S_m$ is called an optimal
noise $D_{opt}$. If the coupling strength is not very strong, $S_m$ takes a
larger value for completely random network ($p=1$) than for regular network
($p=0$). On the other hand, as the coupling strength increases further $S_m$
shows a rapid decline for the more random network. For large $p$ the interaction
among neurons becomes more fast due to long-range connections which amplify the
effect of coupling for the entire range of $g$. Therefore, when the coupling 
is not very strong the long-range connections enhance further the coherent 
motion of the system. However, for the very strong coupling strength, the 
stronger $D_{opt}$ is required due to the reinforced synchronization
by long-range connections. Hence $S_m$ decreases rapidly
for large $p$. The result of Fig. \ref{fig1}(a)-(b) corresponds to the study
of Ref. \cite{aesrGao}. Figure \ref{fig1}(c)-(d) show the results for the cases
of non-zero $\lambda$. The overall tendency does not change except that the
coherent motion is depressed when compared to the case with an uncorrelated
noise. Generally the correlation of noise makes the firing events of neurons
correlated so the effect of mutual excitation is diminished. As a result a
coherent motion is depressed \cite{aecrZhou,crcnWang}.
Interestingly, we found that when $p=0$ and $\lambda=2$, $D_{opt}$
has a relatively low value even for a large coupling strength. When a neuron
is excited by a noise, its neighboring neurons are excited by the coherent
noise most probably. Consequently, spatially correlated noise enables neurons
to generate firing events in the smaller noise intensity even for a relatively
strong coupling. However, because the correlation of noise disappears mostly
again for $p$ near 1, $D_{opt}$s return to those values of the case for $p=1$
and $\lambda=0$.

\begin{figure}
\includegraphics{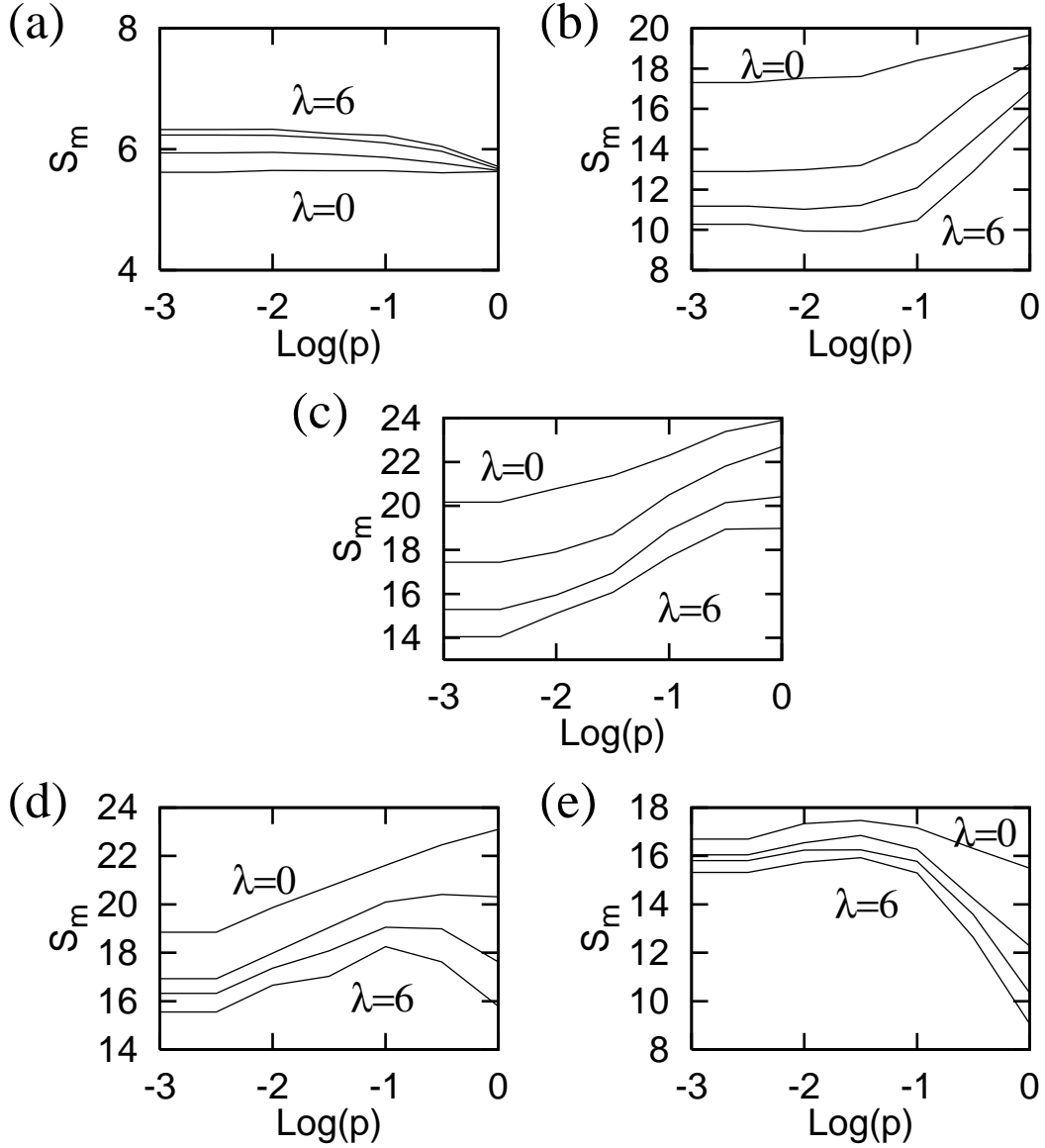}
\caption{\label{fig2} The maximal coherence factor $S_m$ as a function of
network randomness $p$ for different values of $\lambda$,
when coupling strength is (a) very weak $g=10^{-2.5}$, (b) rather weak
$g=10^{-2.25}$, (c) optimal $g=10^{-1.75}$, (d) rather strong $g=10^{-1.25}$,
and (e) very strong $g=10^{-0.75}$.
$\lambda$ is varied from $\lambda=0$ (local noise) to $\lambda=6$ with a step 2.
$S_m$ curve goes down with increasing $\lambda$ except in (a).}
\end{figure}

We then systematically examined the maximal coherence factor $S_m$ as a
function of network randomness $p$ for different values of $\lambda$ (see Fig.
\ref{fig2}). We fix coupling strengths at five values; very weak
($g\approx10^{-2.5}$), rather weak ($g\approx10^{-2.25}$), optimal
($g\approx10^{-1.75}$), rather strong ($g\approx10^{-1.25}$), and very strong
($g\approx10^{-0.75}$). For some cases at $\lambda=0$ (see Fig.
\ref{fig2}(c)-(d)), $S_m$ rises rapidly as $p$ increases when the coupling
strength is not very strong. And $S_m$ rather decreases as $p$ increases for
a stronger coupling $g\approx 10^{-0.75}$ (see Fig. \ref{fig2}(e)). 
The similar behaviors have also been observed in the SR \cite{aesrGao}.

Some interesting features are also found in a certain region of parameter
space. Contrary to an expectation from the existing studies \cite{aecrZhou,
crcnWang}, the spatial correlation of noise enhances a coherent motion. If the
coupling strength is very weak and also if $\lambda$ is zero, each element
behaves independently. However, for $\lambda>0$, $S_m$ curve is elevated
slightly according to $\lambda$ (see Fig. \ref{fig2}(a)) since the spatially
correlated noise assists partially synchronized excitation among the neighboring
neurons even for a very weak coupling. As $p$ increases, the effect of the
correlated noise disappears and the partial synchronization vanishes too.
Therefore, for each value of $\lambda$, $S_m$ decreases with increasing $p$ and
finally reaches the value at $\lambda=0$. For other values of coupling
strength, however, coherence deteriorates in general by the spatial
correlation of noise.

\begin{figure}
\includegraphics{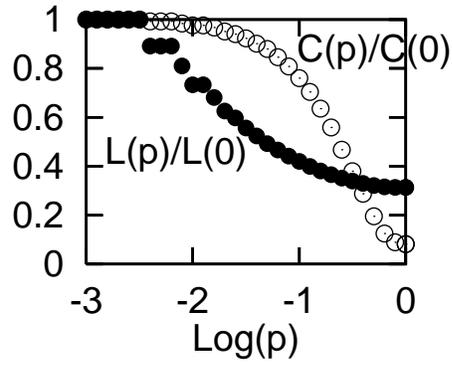}
\caption{\label{fig3} Clustering coefficient $C$ (open circle) and
characteristic path length $L$ (solid circle) as a function of $p$ 
for Watts-Strogatz small-world networks with $N=101$ and $k=6$.
They are normalized by each value at $p=0$.}
\end{figure}

All other $S_m$ curves in Fig. \ref{fig2} are divided roughly into two
categories depending on whether $S_m$ curve monotonically increases or not.
In the case of rather weak coupling the effect of long-range coupling is
clearly observed. When $\lambda$ is large for this case,
we can define the transition point of $p$ around $p_c=0.1$ above which $S_m$
grows significantly. In Fig. \ref{fig3} the clustering coefficient
$C$ remains practically unchanged (clustered structure mostly remains) for
$p<p_c$ while the characteristic path length $L$ drops sufficiently. As $p$
passes over $p_c$ $C$ drops rapidly while $L$ rarely changes.
It indicates that introducing a few long-range connections is enough to
decrease $L$ sufficiently and additional long-range connections for $p>p_c$
affect only to fracture the clustered structure.
Because of the clustered structure of neurons consolidated by a spatially
correlated noise, a few long-range connections do not affect the elevation of
the $S_m$ at large $\lambda$ for a rather weak coupling. As $p$ is increased
beyond $p_c$, the cluster begins to be fractured and then the effect of
noise correlation vanishes considerably. This makes $S_m$, which is definitely
dropped due to the correlated noise at small $p$, grow rapidly to the
value it has when $\lambda=0$ at large $p$. Since the coupling term in Eq.
\ref{eq1} affects the dynamics for a larger coupling strength, the change of
connectivity becomes more crucial. By every rewiring to the long-range
connections the coherent motion is steadily enhanced as $p$ increases for
entire $\lambda$. This ascending behavior of $S_m$ due to the a few long-range
connection is maintained to a rather strong coupling (see Fig.
\ref{fig2}(c) and (d)).

\begin{figure}
\includegraphics{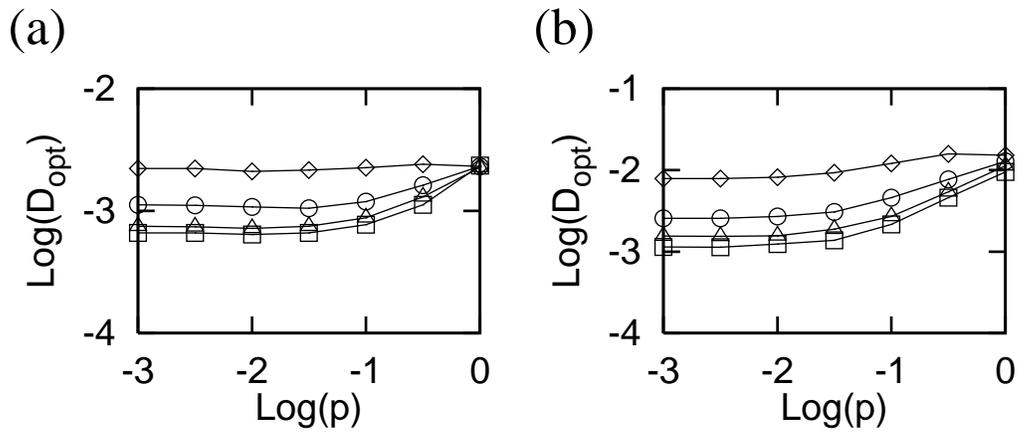}
\caption{\label{fig4} Optimal noise intensity $D_{pot}$ as a function of $p$
for various $\lambda$ including 0 (diamond), 2 (circle), 4 (triangle) and
6 (rectangle). (a) $g=10^{-1.25}$ and (b) $g=10^{-0.75}$.}
\end{figure}

When we consider both the noise correlation and the connection topology, we
observe another peculiar behavior of $S_m$. $S_m$ has a maximum value at an
intermediate $p$ around $p_c$ for a rather strong coupling when $\lambda>0$
(see Fig. \ref{fig2}(d)). In the previous studies $S_m$ either increases or
decreases monotonically \cite{aesrGao, aecrOkyu}. For this to make sense, we
need to follow the behavior of $D_{opt}$ for each situation. When $\lambda=0$
the optimal noise intensity $D_{opt}$ corresponding to $S_m$ rarely changes
for the entire range of $p$ (see Fig. \ref{fig4}(a)). In case of $\lambda>0$,
$D_{opt}$ remains constant at lower level than in case of $\lambda=0$ until
$p\approx p_c$ and after that point gradually chases the value of $\lambda=0$
(totally uncorrelated noise). As has been mentioned above, the
correlated noise enables resonance curve to occur in smaller noise level
especially for relatively strong coupling. This indicates that the effect of
correlation of noise between inter-neurons drastically vanishes as clustered
structure is fractured. As $p$ increases further from $p_c$ for $\lambda>0$,
resonance curve shifts to higher noises. This is why the value of $S_m$ does
not increase and even decreases after $p>p_c$.

Finally for the very strong coupling (see Fig. \ref{fig2}(e)), $S_m$ changes a 
little until $p=p_c$ then decreases substantially for $p>p_c$. This depression
is more distinct as $\lambda$ increases. A few long-range connections do not
affect $S_m$ much, because of the clustered structure consolidated by 
strong coupling and moreover correlated noise especially for large $\lambda$. 
The behavior of $D_{opt}$ as a function of $p$ for various $\lambda$ is very
similar to the case of rather strong coupling as seen in Fig. \ref{fig4}(b).
$S_m$ should decrease much as $\lambda$ increases when $p$ is near zero.
However the decrease is not big owing to the effect of the reduced $D_{opt}$.
This tendency also appears in rather strong coupling case.
As a whole, for $p>p_c$, the movement of $D_{opt}$ toward larger value
and a number of long-range connections with very strong coupling drastically
reduce the temporal coherence of the system.

%Conclusion
In summary, we have investigated the effect of spatially correlated noise
(correlation length $\lambda$) in the presence of various connection topology
(network randomness parameter $0<p<1$) of FHN neural network. This study
reproduces most of the general feature of AESR and AECR in the entire range
of $\lambda$ and $p$. In addition, we could get some novel features for
$\lambda>0$. When the coupling is rather strong, an optimum value
$p \approx p_c$ clearly emerges where a maximal coherence resonance appears.
For $p$ increasing from 0 to $p_c$, the maximal coherence factor rarely
changes but, beyond $p \approx p_c$, it grows dramatically either up for a weak
coupling or down for a strong coupling. For $p$ beyond $p_c$, it is observed
that the clustered structures of the neurons are mostly fractured out due to
many long-range connections and, as a result, the noise correlation of
inter-neurons is diminished quickly. It is believed to be the reason for such
an abrupt change of $S_m$ near $p \approx p_c$. These results show that the
spatially correlated noise enforces the role of the clustered structure to the
system. Therefore, the effect of a few long-range connections is ignored by
the enhanced clusters of the neurons for a weak as well as for a strong
coupling. Nevertheless, for an optimal and for a rather strong coupling,
coherence resonance is still considerably enhanced by a small portion of
long-range connections.

We thank O. Y. Kwon for the computer resource for the numerical computations.

% bibliography
\bibliography{scn_on_cr}

\newpage

\end{document}